\documentclass[final,5p,times,twocolumn]{elsarticle}
\pdfoutput=1

\usepackage{graphicx}

\usepackage{amssymb}

\usepackage{hyperref}

\journal{Nuclear Instruments and Methods A}

\begin{document}

\begin{frontmatter}

\title{Neutral bremsstrahlung in two-phase argon electroluminescence: further studies and possible applications}

\author[A,B]{A. Bondar}
\author[A,B]{A. Buzulutskov}
\author[B]{A. Dolgov}
\author[A,B]{E. Frolov}
\author[A,B]{V. Nosov}
\author[A,B]{V. Oleynikov}
\author[A,B]{E. Shemyakina\corref{corresponding author}}
\ead{E.O.Shemyakina@inp.nsk.su}
\author[A,B]{A. Sokolov}
\address[A]{Budker Institute of Nuclear Physics SB RAS, Lavrentiev avenue 11, 630090 Novosibirsk, Russia}
\address[B]{Novosibirsk State University, Pirogov street 2, 630090 Novosibirsk, Russia}

\cortext[corresponding author]{Corresponding author}






\begin{abstract}
We further study the effect of neutral bremsstrahlung (NBrS) in two-phase argon electroluminescence (EL), revealed recently in \cite{NBrSEL18}. The absolute EL yield due to NBrS effect, in the visible and NIR range, was remeasured in pure gaseous argon in the two-phase mode, using a two-phase detector with EL gap read out directly by cryogenic PMTs and SiPMs. Possible applications of the NBrS effect in detection science are discussed, including those in two-phase dark matter detectors.
\end{abstract}


\end{frontmatter}


\section{Introduction}

The effect of proportional electroluminescence (EL) in noble gases \cite{ArELTheory11,ArXeN2Rev17} has long been used in two-phase detectors to record ionization signals in the gas phase, induced by particle scattering in the liquid phase  (so-called S2 signals) \cite{Chepel13,DarkSide20k18}. Such two-phase detectors are relevant for dark matter search and low-energy neutrino detection.  The S2 signals are recorded typically in the EL gap placed above the liquid-gas interface, optically read out by either cryogenic PMTs or cryogenic SiPMs. 

Until recently it was believed that proportional electroluminescence was fully due to vacuum ultraviolet (VUV) emission of noble gas excimers, e.g. Ar$^{\ast}_2(^{1,3}\Sigma^+_u)$, produced in three-body atomic collisions with excited atoms, e.g.  Ar$^{\ast}(3p^54s^1)$, which in turn are produced by drifting electrons in electron-atom collisions: see review \cite{ArXeN2Rev17}. In the case of Ar this results in almost mandatory use of a wavelength shifter (WLS) in front of PMTs and SiPMs, to convert the VUV emission into the visible light \cite{DarkSide20k18}. 

On the other hand, an additional mechanism of proportional electroluminescence has been recently revealed \cite{NBrSEL18}, namely that of bremsstrahlung of drifting electrons scattered on neutral atoms (so-called neutral bremsstrahlung, NBrS). It was stated that the NBrS effect can explain two intriguing observations in EL radiation: that of the photon emission at lower electric fields, below the Ar excitation threshold, and that of the noticeable contribution of the non-VUV spectral component, extending from the UV to NIR. The latter may pave the way for direct optical readout of two-phase Ar detectors,  without using WLS. 

In this work, we further study the  NBrS effect in proportional electroluminescence in two-phase Ar. In particular, the absolute EL yield in pure Ar is remeasured  using a dedicated two-phase detector with EL gap, read out directly (without WLS) by cryogenic PMTs and SiPMs. We also discuss the possible applications of the NBrS effect  in two-phase dark matter detectors.

\section{Experimental setup and procedures}

\begin{figure}[htb]
	\centering
	\includegraphics[width=0.99\columnwidth,keepaspectratio]{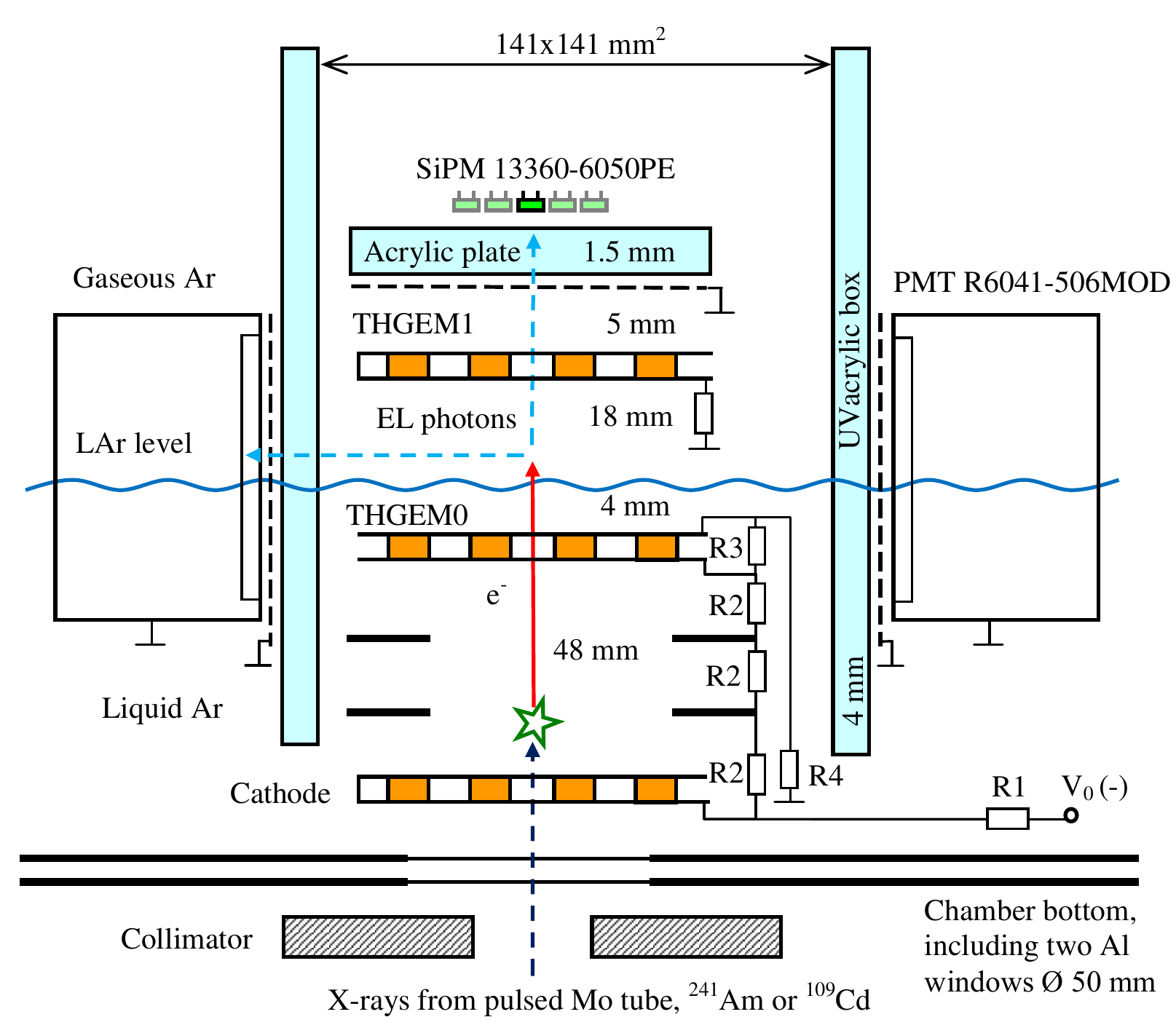}
	\caption{Schematic view of the experimental setup (not to scale).}
	\label{Setup}
\end{figure}

Fig.~\ref{Setup} shows the experimental setup for the present measurement session, of 2019; it was modified compared to the previous measurement session, of 2018, described in \cite{NBrSEL18} (see below). The setup comprised a two-phase TPC with EL gap, filled with liquid Ar and operated in the two-phase mode at a saturated vapor pressure of 1.0 atm and temperature of 87.3 K.
The EL gap, formed by the liquid surface and THGEM1 electrode, was viewed by four compact PMTs R6041-506MOD \cite{CryoPMT17}, located on the perimeter of the gap. The PMTs were electrically insulated from the gap by an acrylic box. 

An important modification of the setup compared to that of \cite{NBrSEL18} was that all the four PMTs  were bare, i.e. were insensitive to the VUV: in place of the acrylic box with WLS films, a box without WLS was used, made from the UV acrylic. Accordingly, the overall spectral sensitivity of the PMTs ranges from 300 to 600 nm \cite{NBrSEL18}. In addition, doing without WLS results in suppression of non-VUV crosstalk between the PMTs, which otherwise is induced by re-emission of ordinary VUV electroluminescence in the WLS. In this respect, the present experiment can be considered as more "pure" compared to that of \cite{NBrSEL18}.   

The EL gap was also viewed by a 5x5 SiPM-matrix from the top, through an acrylic plate and a THGEM1 electrode (acting as a mask with 27\%  optical transmission), with the overall spectral sensitivity ranging from 400 nm to 1000 nm \cite{NBrSEL18}. Only the central SiPM, of 6x6 mm$^2$ area and 13360-6050PE type \cite{Hamamatsu}, was used in the present work; it was operated at overvoltage of 3.6 V. 


Other details of the experimental setup and measurement procedures were described elsewhere \cite{NBrSEL18}.

\section{Results}

Compared to previous work \cite{NBrSEL18}, we modified the method to  associate the PMT and SiPM signal amplitude with the photoelectron (p.e.) number. In the previous method, described elsewhere \cite{CRADPropEL17}, the PMT and SiPM p.e. numbers were determined dividing the signal pulse area by that of the single-electron pulse. In the present work we directly counted the number of p.e. peaks in the pulse waveforms, using a peak finder algorithm. The old method may have uncertainties due to ignoring crosstalk and baseline shift, while the new one due to peak overlapping. To reduce these and other systematic uncertainties, in what follows we present the data averaged over the two measurement sessions, of 2018 and 2019. 

\begin{figure}[!hbt]
	\centering
	\includegraphics[width=0.99\columnwidth,keepaspectratio]{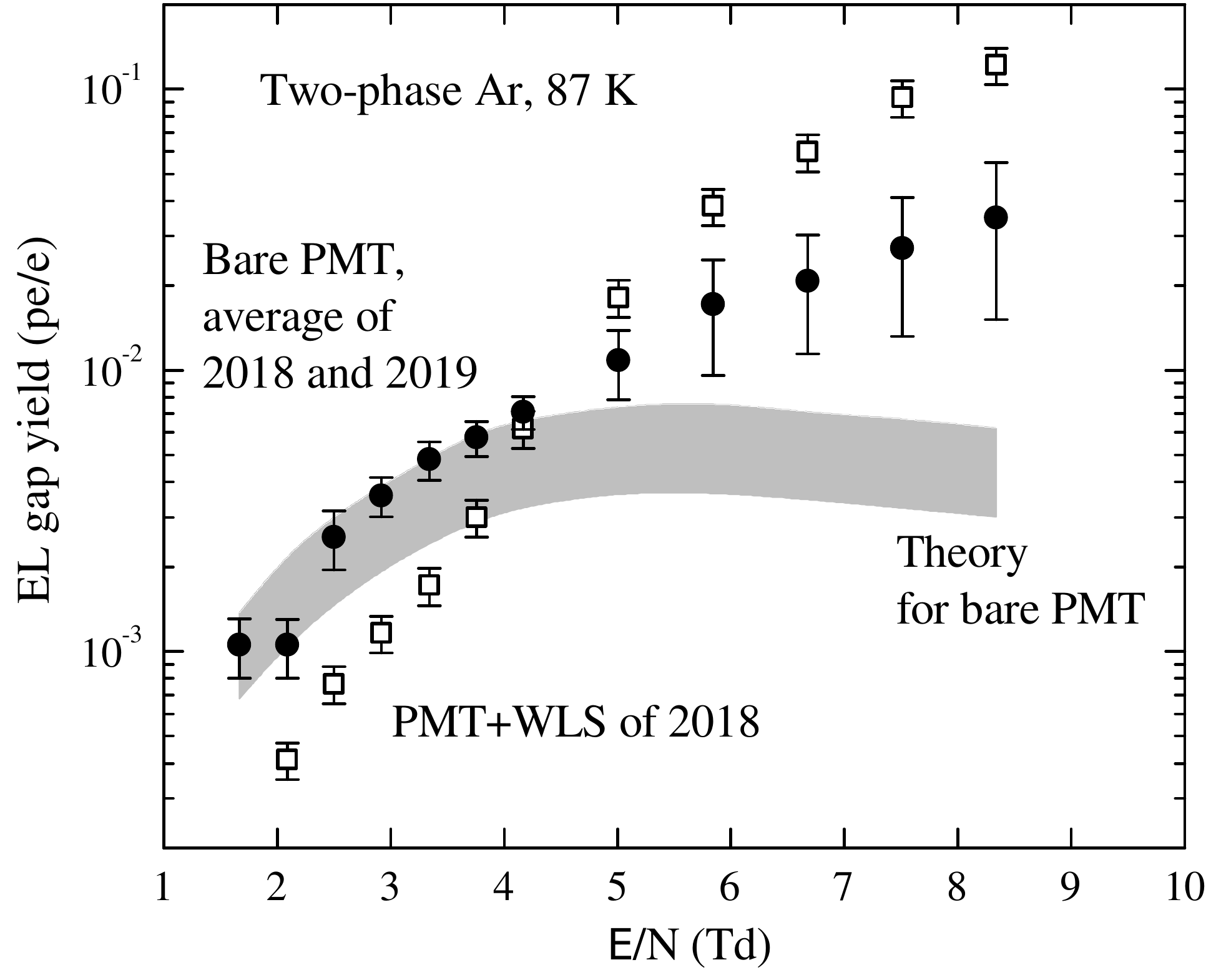}
	\caption{EL gap yield for a single bare PMT readout (average of 2018 and 2019 measurement sessions, closed data points) and that of PMT+WLS (2018 measurement session, open data points) as a function of the reduced electric field. The prediction of the theory of NBrS electroluminescence for the bare PMT readout \cite{NBrSEL18} is shown by the grey area.}
	\label{EL-gap-yield-PMT}
\end{figure}

\begin{figure}[!hbt]
	\centering
	\includegraphics[width=0.99\columnwidth,keepaspectratio]{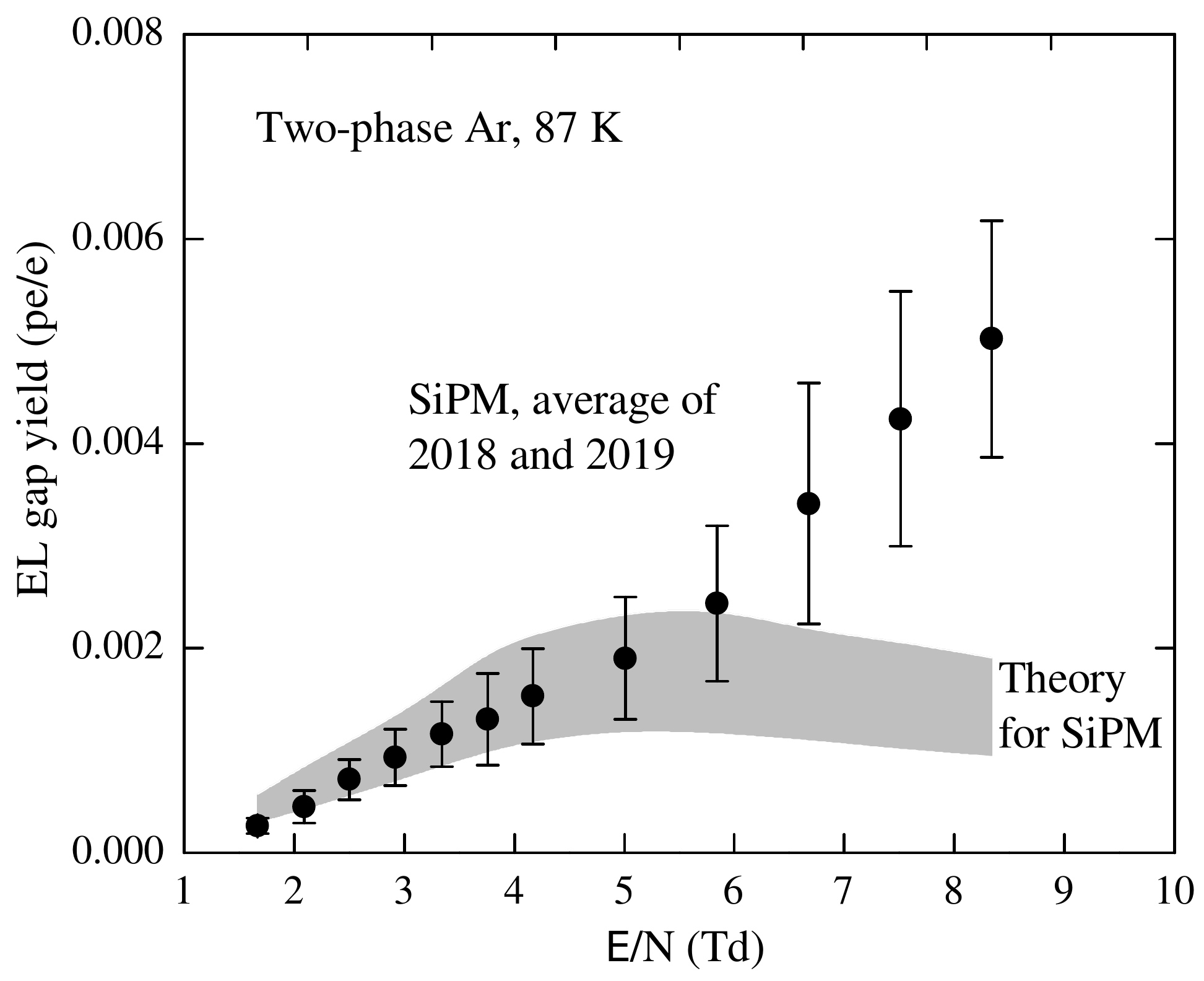}
	\caption{EL gap yield for the central SiPM readout at an overvoltage of 3.6 V (average of 2018 and 2019 measurement sessions) as a function of the reduced electric field. The prediction of the theory of NBrS electroluminescence for SiPM readout \cite{NBrSEL18} is shown by the grey area.}
	\label{EL-gap-yield-SiPM}
\end{figure}

Fig.~\ref{EL-gap-yield-PMT} shows the EL gap yield as a function of the reduced electric field for the bare PMT readout and that of PMT+WLS, along with the prediction of the theory of NBrS electroluminescence \cite{NBrSEL18}. Similarly, in Fig.~\ref{EL-gap-yield-SiPM} the experimental and theoretical EL gap yields are compared for the central SiPM readout. Note that the reduced electric field of 1 Td = 10$^{-17}$ V cm$^2$, corresponding to 0.87 kV/cm in gaseous Ar at 87.3 K.

One can see that all principle results of the previous work \cite{NBrSEL18} are confirmed: the noticeable contribution of the non-VUV spectral component in EL radiation, extending from the UV to NIR, and the photon emission at lower electric fields, below the Ar excitation threshold where the non-VUV component fully dominates. 

It is also confirmed that while below the Ar excitation threshold, at 4.0 Td, the non-VUV component is well described by the NBrS theory, above the threshold the theory quickly diverges from the experiment. In \cite{NBrSEL18} it was proposed that such a discrepancy might be explained by the effect of sub-excitation Feshbach resonances \cite{Schulz73}, which may enhance the intensity of the NBrS emission \cite{Dyachkov74}. Relying on these hypotheses and similarly to \cite{NBrSEL18}, we adopt the NBrS paradigm stating that all the data on the non-VUV component in proportional electroluminescence are those due to NBrS mechanism. 

Within this paradigm we calculated the absolute photon yields both for ordinary electroluminescence (in the VUV) and for that of NBrS (for wavelengths not exceeding 1000 nm), using experimental data of Figs.~\ref{EL-gap-yield-PMT} and \ref{EL-gap-yield-SiPM} and NBrS emission spectra at a given electric field \cite{NBrSEL18}. The detailed procedure was described in  \cite{NBrSEL18}. To reduce the systematic errors, the results were averaged over the bare PMT and SiPM data. In particular to obtain the true yield in the VUV, due to ordinary electroluminescence, we subtracted from the PMT+WLS data of Fig.~\ref{EL-gap-yield-PMT} the non-VUV contribution due to the NBrS effect. The latter was calculated using the NBrS emission spectra and the bare PMT and SiPM data. 

\begin{figure}[!hbt]
	\centering
	\includegraphics[width=0.99\columnwidth,keepaspectratio]{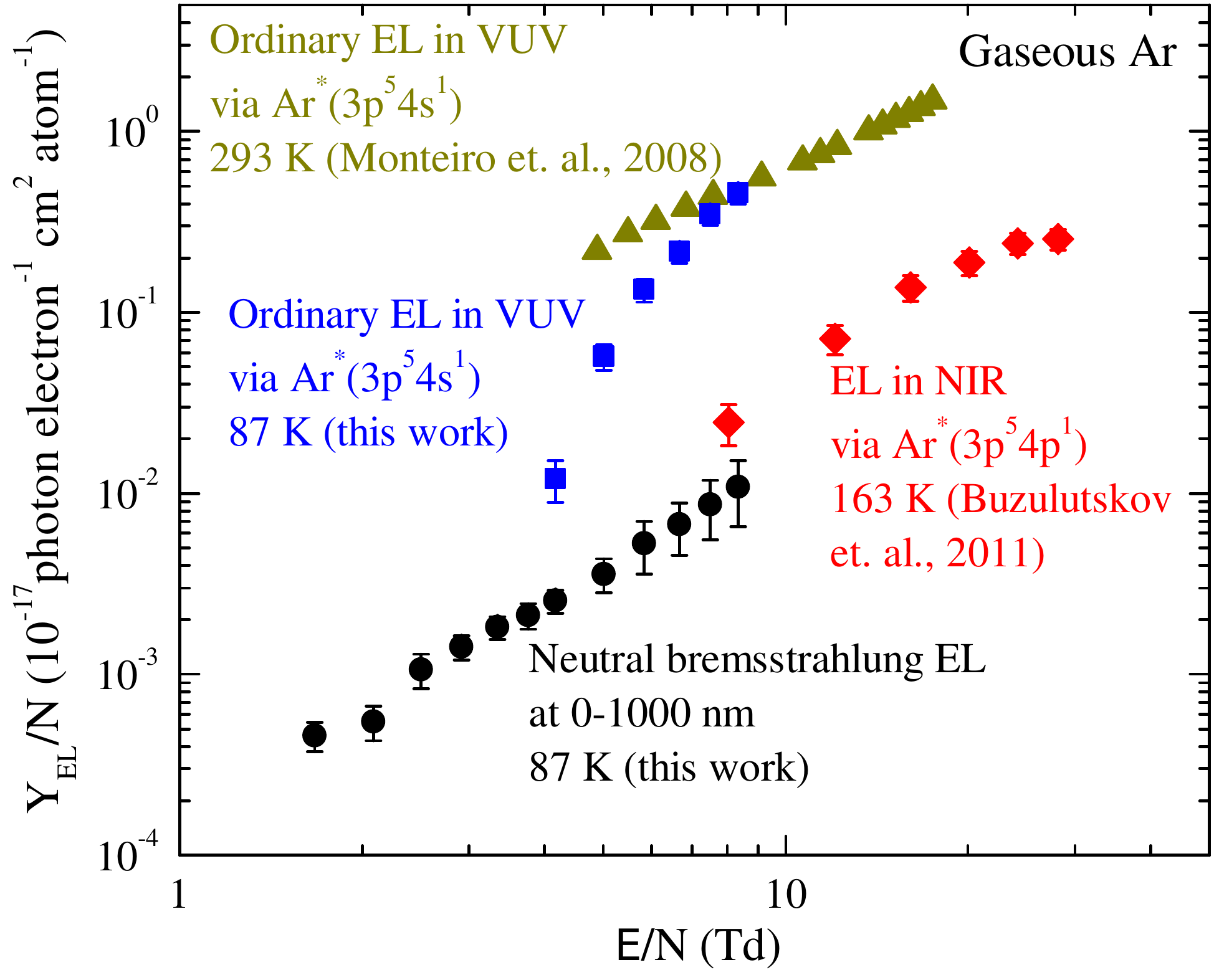}
	\caption{Summary of experimental data on reduced EL yield in argon for all known electroluminescence (EL) mechanisms: for NBrS EL at wavelengths of 0-1000 nm obtained in this work at 87 K from the bare PMT and SiPM data  and using theoretical spectra of NBrS EL emission \cite{NBrSEL18}; for ordinary EL in the VUV, going via Ar$^{\ast}(3p^54s^1)$ excited states, obtained in this work at 87 K, subtracting the non-VUV contribution within the NBrS EL paradigm, and in Monteiro et al., 2008 at 293 K \cite{ArELExp08}; for EL in the NIR, going via Ar$^{\ast}(3p^54p^1)$ excited sates, obtained in Buzulutskov et al., 2011 at 163 K \cite{NIREL11}.}
	\label{EL-yield-summary}
\end{figure}

Summarizing, Fig.~\ref{EL-yield-summary} presents all known experimental data on reduced EL yields in Ar for all known EL mechanisms (\cite{NBrSEL18,ArELTheory11,ArXeN2Rev17}): for NBrS electroluminescence  at wavelengths below 1000 nm, measured in this work at 87 K; for ordinary electroluminescence in the VUV, going via Ar$^{\ast}(3p^54s^1)$ excited states, measured in this work at 87 K and in \cite{ArELExp08} at room temperature; for electroluminescence in the near infrared (NIR) going via Ar$^{\ast}(3p^54p^1)$ excited states, measured in \cite{NIREL11} at 163 K. Note that for ordinary electroluminescence the data of \cite{ArELExp08} have an excess over our data at fields below 6 Td. It would be logical to explain this discrepancy by the NBrS contribution that was not taken into account in \cite{ArELExp08}. 

One can see that NBrS electroluminescence is the weakest among all mechanisms. On the other hand, it exists in the whole range of electric fields, without having a threshold, in contrast to VUV (ordinary) and NIR electroluminescence.  The two latter goes via excited atomic states and thus have thresholds in electric fields, at about 4 and 7 Td respectively.

\section{Possible applications of NBrS luminescence}

One can see from Fig.~\ref{EL-yield-summary} that NBrS electroluminescence cannot compete with ordinary and NIR electroluminescence in terms of intensity for higher electric fields, above 7 Td. However, below this value, where NIR electroluminescence disappears and VUV electroluminescence is not that strong, the response of PMT and SiPM to NBrS electroluminescence may be comparable with the response of PMT+WLS to ordinary electroluminescence. This is because in the absence of optical contact between the WLS and the PMT the photon flux is considerably  reduced after re-emission by the WLS, by about a factor of 15-20 \cite{CRADPropEL17}, due to re-emission and total reflection losses.

\begin{figure}[!hbt]
	\centering
	\includegraphics[width=0.99\columnwidth,keepaspectratio]{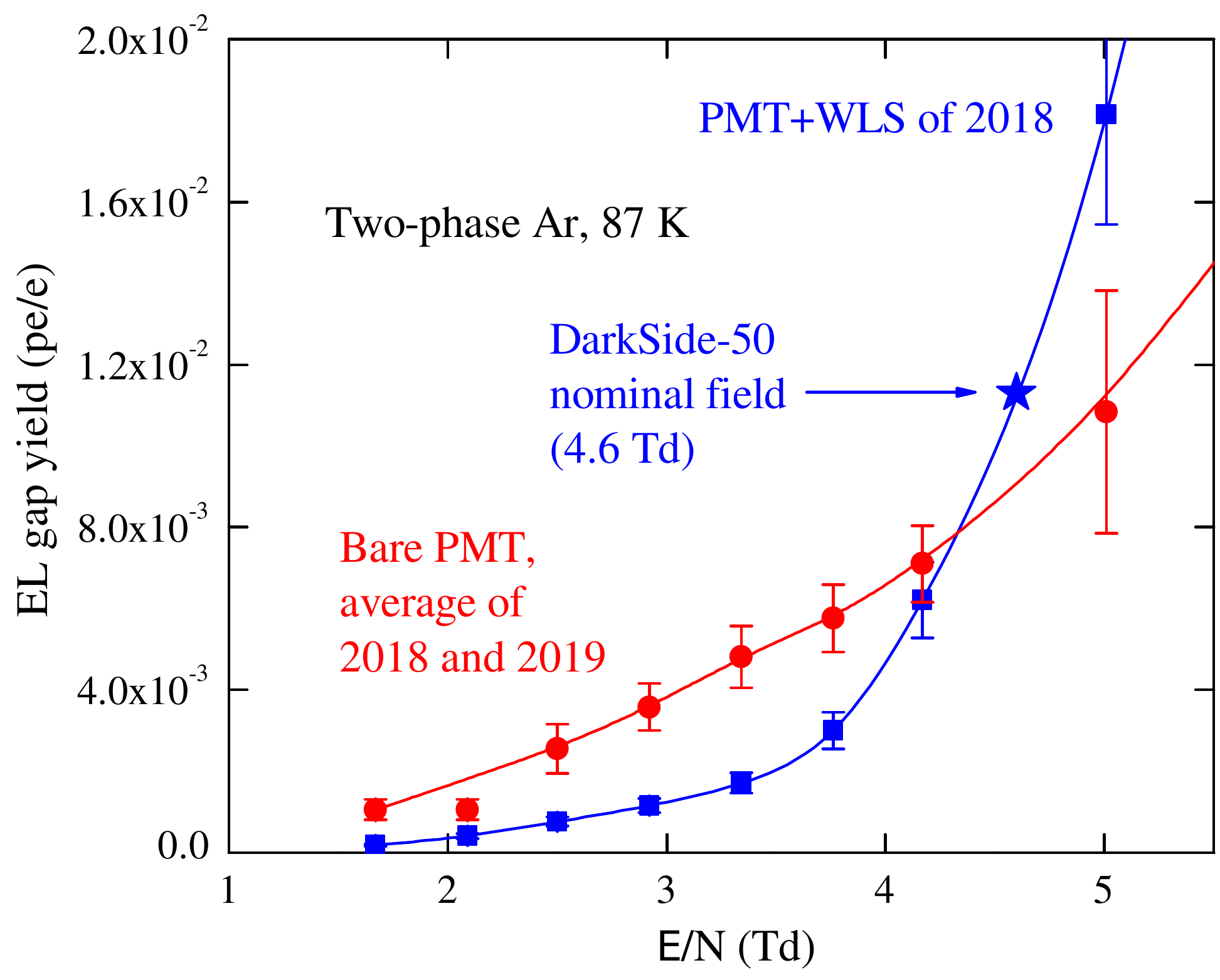}
	\caption{EL gap yield at moderate electric fields for PMT+WLS (of 2018 measurement session) and bare PMT (average of 2018 and 2019 measurement sessions). For PMT+WLS, the EL gap yield at DarkSide-50 nominal electric field \cite{DarkSide15} is indicated by a star.}
	\label{EL-gap-yield-PMT-lowfield}
\end{figure} 

Accordingly, the important conclusion is that at moderate electric fields in the EL gap, below 5 Td, the amplitude of the S2 signal  from the bare PMT might be comparable with that of the PMT with WLS: see Fig.~\ref{EL-gap-yield-PMT-lowfield}. This result is particularly relevant to the DarkSide-50 dark matter search experiment \cite{DarkSide15} where the bare PMT response might be almost the same as that of PMT+WLS at the nominal operating field, of 4.6 Td. 

The S2 amplitude can be further increased, by a factor of 2, if to replace the bare PMTs with the SiPM-matrices, since the latter have higher PDE and wider range of sensitivity to NBrS spectra \cite{NBrSEL18}. This observation paves the way for direct readout of S2 signals in two-phase dark matter detectors: via NBrS electroluminescence in the visible and NIR range using PMTs and SiPM-matrices. 

In particular, such a direct readout of a two-phase Ar detector in the visible and NIR range, using a 5x5 SiPM-matrix, has been recently demonstrated by our group \cite{SiPM-matrix-readout}. For 80 keV gamma-rays, the SiPM-matrix yield amounted to 0.4 p.e. per keV of deposited energy in the present (not-optimized) readout configuration shown in Fig.~\ref{Setup}. For optimized configuration it is expected to increase by an order of magnitude. 

The presence of the NBrS component in proportional electroluminescence, which is naturally fast, may result in suggesting to analyze the S2 pulse-shape in a new way, in particular in two-phase Ar dark matter detectors \cite{DS-S2-PulseShape}. For example, at 4.6 Td (at DarkSide-50 operating field) one has to take into account the substantial enhancement of the fast component due to NBrS electroluminescence. Indeed, the NBrS contribution to the total S2 signal recorded by PMTs with WLS is estimated to be about 50\% \cite{NBrSEL18}. Such an enhancement can affect the algorithm for decomposing the S2 signal into the fast and slow component, which in turn can affect the determination of the quantities using the fast component, such as the diffusion coefficients in liquid Ar \cite{DS-S2-PulseShape} or z-coordinate fiducialization. 

It should be emphasized that NBrS electroluminescence has a universal character: since its intensity is proportional to elastic cross section in electron-atom collisions \cite{NBrSEL18}, it should be present in the gases dominated by elastic scattering of electrons, i.e. in all noble gases. That is why we assume that NBrS electroluminescence is present in S2 signals of two-phase Xe detectors. Presumably it has not been yet observed due to the fact that the S2 signal in Xe is recorded directly using PMTs with quartz windows (i.e. unlike Ar, without losses due to re-emission in WLS), and at higher electric fields, to provide efficient extraction of the electrons from liquid Xe. This left almost no chance for NBrS signal to be observed at the background of a strong main signal. We expect that NBrs electroluminescence in Xe will be observed whenever the VUV component is suppressed (for example by optical filters). 

It is worth mention the possible application of NBrS radiation  to develop a detection technique for ultra-high-energy cosmic rays \cite{AlSamarai16}. Here the NBrS radiation in the radio-frequency range  (to be recored with antennas
on the Earth surface) is emitted by primary ionization electrons left after the passage of the showers in the atmosphere. 

It was suggested \cite{NBrSEL18} that the similar phenomenon could be responsible for the weak primary scintillations in liquid Ar (S1 signals) in the visible and NIR range, observed earlier by a number of groups \cite{NIREL11,Heindl10,Alexander16}: such primary scintillations in the non-VUV might be explained by neutral bremsstrahlung of the primary ionization electrons, decelerated in the medium down to the energy domain of elastic electron-atom collisions.

It was also supposed \cite{NBrSEL18} that the NBrS effect can be responsible for proportional electroluminescence observed in liquid Ar and Xe using immersed GEM-like structures \cite{Lightfoot09,CRADRev12}. Indeed, the electric fields in the center of GEM or THGEM holes used in liquid Ar, of 60-140 kV/cm \cite{CRADRev12}, correspond to $\mathcal{E}/N=0.3-0.7$~Td. For such reduced electric fields the theory predicts that NBrS electroluminescence already exists \cite{NBrSEL18}. 

Finally, the NBrS effect has long been known in plasma physics: it was used to explain continuous emission spectra in arc \cite{Batenin72}, glow \cite{Rutscher76} and RF \cite{Park00} discharges. It was also used to explain emission spectra in sonoluminescence \cite{Frommhold98}.

\section{Conclusions}

In this work we further studied the neutral bremsstrahlung (NBrS) mechanism of proportional electroluminescence (EL) in gaseous Ar in the two-phase mode, revealed in \cite{NBrSEL18}. All principle results of the previous work has been confirmed: the noticeable contribution of the non-VUV spectral component in EL radiation, extending from the UV to NIR, and the photon emission at lower electric fields, below the Ar excitation threshold. 

It is also confirmed that while below the Ar excitation threshold the non-VUV component is well described by the NBrS theory, above the threshold the experiment quickly diverges from the theory. We expect that such a discrepancy will be explained by the effect of sub-excitation Feshbach resonances: the theoretcial and experimental studies in this direction are in the course in our laboratory.

The NBrS effect has a universal character: it should be present in all noble and molecular gases. It may also explain the non-VUV components observed earlier in various light emission processes, in particular the primary and secondary scintillations in noble liquids in the visible and NIR range.

The main practical application of the NBrS effect is a better understanding of the S2 signal, in particular of its time structure, and justification for its direct (without WLS) optical readout using PMTs and SiPM-matrices. This may help to develop the robust and ultrasensitive two-phase detectors for dark matter search and low energy neutrino detection.

\section*{Acknowledgments}

This work was supported by Russian Foundation for Basic Research (project no. 18-02-00117). It was done within the R\&D program of the DarkSide-20k experiment. 



\end{document}